# SDN Based QoS Provision in WSN Technologies

Babedi. B. Letswamotse[1], Kgotlaetsile. M. Modieginyane[2] and Reza Malekian[3]

[1, 2, 3]Department of Electrical, Electronic and Computer Engineering, University of Pretoria, Private bag X20, Hatfield 0028, South Africa

[1]bletswamotse@gmail.com, [2]lekgotla.magivo@gmail.com, [3]reza.malekian@up.ac.za

*Abstract—* **Wireless Sensor Networks (WSNs) have rapidly progressed over the years, they are now applied in health care systems, home automation, security surveillance, disaster management and more. With this rapid growth in the number of users and applications leading to WSNs becoming increasingly complex, this growth makes high demands on WSNs to provide the requirements of every user and application that uses them. They have recently been envisioned to be integrated into Internet of Things (IoT), their role will be to provide sensing services to the ever increasing community of internet users. However even with so much potential, WSNs still experience issues in node deployment, fault tolerance, scalability and Quality of Service (QoS) provisioning amongst others. In this paper we propose to improve QoS provisioning by introducing Software Defined principles into WSN technologies.**

## I. INTRODUCTION

Over the past years WSNs have become an extremely dynamic and most fascinating field of research in both the industrial and academic territories, due to their potential functions and applications in health care systems, motion detection, military surveillance, environmental observation and traffic control to name a few. Of late they have been regarded as the principal elements in the IoT paradigm. IoT is a network in which both wired and wireless everyday objects like internet TV and smart devices that have internet browser and wireless access are interconnected to the World Wide Web, where these objects communicate and provide information and knowledge to other objects and people[1].

WSNs are widespread networks consisting of groups of specialized, relatively small network nodes with the communication infrastructure (sensing circuitry, an approximate amount of memory, a microcontroller, a wireless transceiver and a power source) for monitoring and recording conditions at diverse locations. These nodes are expected to join forces to perform application specific task that are ranging from object, animals and human tracking, intrusion detection, disaster management and more. Even with such a great potential, WSNs are severely resource constrained with regards to memory, processing capacities, communication capabilities and energy [2].

Our research focus is on the provision of the overall desired QoS by introducing Software Defined Networking (SDN) in WSNs. In a WSN, QoS is the ability of the network to handle traffic in a manner that it is able to provide the required preferential service delivery by ensuring high quality performance in terms of guaranteeing sufficient bandwidth, controlling latency and jitter and reducing data loss [3]. QoS metrics usually include: bandwidth, throughput, latency, delay, jitter, packet loss and coverage. However recently energy consumption and network lifetime have become standard part of QoS metrics [4]. SDN is a paradigm that allows network administrators to manage network services through abstraction of lower level functionalities. The basic idea of SDN is to separate the data plane from the control plane in network devices and to centralize control of the network operation [5].

## II. MOTIVATION

With the notion that due to increasing number of users and applications, soon it will not be sufficient to have QoS only as the option in WSNs, rather QoS must become a standard part of WSNs. It is of paramount importance to provide the overall desired QoS guarantee. However, maintaining the desired QoS provisioning is not easy because WSNs go through changes recurrently, the changes may be due to the network topology change, load imbalance, faulty nodes and unstable wireless links. Consequently it is important to develop QoS techniques that are dynamic, energy saving and resilient to frequent changes. Accordingly this research proposes the use of SDN in WSNs as a means to improve the QoS provision in WSNs.

## III. LITERATURE

Over the years researchers have worked extensively to provide the desired QoS requirements for WSNs and they came up with various methodologies ranging from QoS aware routing protocols layered and cross layered QoS support approaches and middleware layer QoS support. In [6] System Communication Services (SCS) protocol for hybrid WSNs to optimize QoS provisioning was proposed, SCS was compared against DSR and AODV protocols in terms of average delay, throughput and packet delivery ratio. Performance results reveal that unlike AODV, SCS performance increases regardless of the fact that the number of malicious nodes is increased, moreover SCS performs much better than AODV and DSR protocols. The other efforts include: AODV and OLSR [7], EQSR [8], PQMAC [9] and QoSNET [10].

An investigation by Nefzi et al. [11] reported that the SCSP (Server Cache Synchronization Protocol) standard protocol, provides remarkable benefits in terms of cooperating with the MAC layer to enhance its operational mechanisms for fault tolerance. Furthermore, this work has also indicated that this protocol does not require routing maintenance as routing paths could be selected based on the collision information recorded on the MAC.

The work of Cordeiro et al. in [12] anticipated the approach of energy-efficient cross-layer design on the network layer and the MAC layer as aim to lessen the overall energy consumption within the network.

However these methodologies do not provide the overall QoS support. Some of the techniques meet a certain level of QoS but at the expense of the energy. The layered approach meet QoS requirements on individual layers. Cross layer approach seems to have a lot of potential but most of the proposed cross layer approaches are not implemented in real networks and therefore lack real world experimental proof.

## IV. PROPOSED RESEARCH

Our work aims to provide the overall QoS without breaking down the communication protocol stack. We propose to implement a QoS support technique that is energy efficient, resource and data traffic aware and help in prolonging the network lifetime.

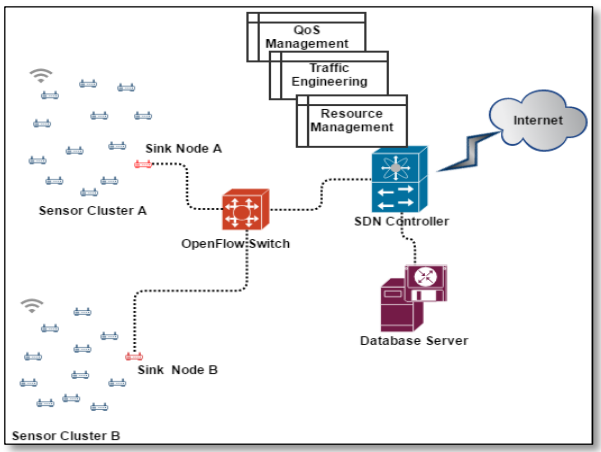

**Fig1**: Software Defined WSN (SDWSN) proposed Architecture

A. We will extensively review the previous work done on QoS, resource management and SDN to investigate how SDN based QoS support can be guaranteed without compromising energy usage and develop a general QoS framework.
B. We will design and model the SDWSN structure as shown in **Fig1** by simulation and emulation using Omnet++ and MiniNet. Where we will:
  1) Develop an algorithm that will guarantee efficient resource allocation and utilization.
  2) Develop a function that will check the status of all the network resources.
  3) Develop an intelligent technique for prolonging Network Lifetime for SDWSNs.
  4) Implement an OpenFlow based energy saving QoS mechanism or technique for SDWSNs.
C. Upon successful completion of our work (simulations and emulations) we will implement our final solution on an already existing network and compare with other current existing methods used to improve the provisioning and ensuring of QoS in WSNs to test its feasibility as proof of concept.

## V. CONCLUSION

WSNs have become a part of our lives and it is of great significant to ensure the QoS of WSN technologies. Our work aims to take advantage of the SDN benefits and apply them in WSNs to help improve QoS provisioning. On successful completion of our work, we will be able to provide better QoS guarantee that is traffic, resources and energy aware.

### ACKNOWLEDGMENT

The Authors would like to thank The National Research Foundation and The Telkom Centre of Excellence for their support.

## REFERENCES

[1] R. Piyare and S. Ro Lee, "Towards Internet of Things (IoTs): Integration of Wireless Sensor Network to Cloud Services for Data Collection and Sharing." International Journal of Computer Networks & Communications (IJCNC), Vol.5, No.5, pp. 59-72, 2013.
[2] M. Kabiri and J. Vahidi. "Analysis of Performance Improvement in Wireless Sensor Networks Based on Heuristic Algorithms Along with Soft Computing Approach". Journal of mathematics and computer Science, vol. 13, pp. 47-67, Sept. 2014.
[3] I. S. Kocher, C. Chow, H. Ishii, and T. A. Zia. "Threat Models and Security Issues in Wireless Sensor Networks". International Journal of Computer Theory and Engineering, Vol. 5, No. 5, pp. 830-835, 2013.
[4] S. Vural, Y. Tian and E. Ekici In: A. Boukerche, Eds. QoS-Based Communication Protocols in Wireless Sensor Networks. New York: John Wiley & Sons 2008: pp. 365-375.
[5] M. Jammal, T. Singh, A. Shami, R. Asal and Y. Li. "Software-Defined Networking: State of the Art and Research Challenges". Journal of Computer Networks, vol. 72, pp. 74-98, Oct. 2014.
[6] G. Ramprabu, S. Ananthi, R. Chitra, J. Saranya, G. S. Preethi. "Design of SCS Protocol and Analysis of Quality of Service Parameters for Wireless Sensor Networks." International Journal of Innovative Research in Computer and Communication Engineering, Vol. 3, No 3, pg. 1752-1756, Mar. 2015
[7] R. Malekian and A. Karadimce, "AODV and OLSR routing protocols in MANET." In proc. of 2013 IEEE 33rd International Conference on Distributed Computing Systems Workshops, pp.286-289, Jul. 2013, Philadelphia, USA.
[8] B. Yahya and J. Ben-Othman, "A Energy Efficient QoS Aware Multipath Routing Protocol for Wireless Sensor Networks," LNC 2009 IEEE 34th Conference on local Computer Networks, 2009, pp. 93-100.
[9] H. Kim and S.G Min, "Priority-based QoS MAC protocol for wireless Sensor networks" IPDPS 2009, IEEE International Symposium on Parallel and Distributed processing, pp. 1-8, 2009.
[10] T. Houngbadji and S. Pierre, "QoSNET: An Integrated QoS Network for Routing protocols in Large Scale Wireless Sensor Networks, "Computer Communications, Vol. 33, No. 11, pp. 1334-1342, Jul. 2010.
[11] B. Nefzi, H. Cruz and Y.Q. Song, " SCSP: An Energy Efficient Network- MAC Cross-Layer Design for Wireless Sensor Networks," The 9th IEEE International workshop on wireless local networks, 2009 pp.1061-1068.
[12] C.M. Cordeiro, D. P. Agrawal, C.C. Weng, C.W. Chen, P.Y. Chen and K.C. Chang, "Design of an energy-efficient cross-layer protocol for mobile ad hoc networks," IET Communications; Vol. 7, No. 3, pp. 217–228, 2013.

**Babedi Betty Letswamotse** received her undergraduate degree (Computer Science and Mathematics) in 2011, her Honours degree (Computer Science) in 2012 and Master of Science degree (Computer Science) in 2014, from the North West University (Mafikeng Campus) and is currently studying towards her PhD in Electronic Engineering at the University of Pretoria. Her research interests include: GPU general purpose computing, Automatic Speech Recognition, Wireless Sensor Networks and Software Defined Networking.

**Reza Malekian** is currently a Senior Lecturer with the Department of Electrical, Electronic, and Computer Engineering, University of Pretoria, South Africa. His current research interests include: advanced sensor networks, Internet of Things, and mobile communications.